\documentclass[12pt]{article} 
\usepackage{graphicx}
\usepackage{amsmath}
\usepackage{amssymb}
\usepackage{cite}

\makeatletter
 
  \@addtoreset{equation}{section}
 \makeatother

\newcommand{\be}{\begin{equation}}
\newcommand{\ee}{\end{equation}}
\newcommand{\bea}{\begin{eqnarray}}
\newcommand{\eea}{\end{eqnarray}}
\newcommand{\beann}{\begin{eqnarray*}}
\newcommand{\eeann}{\end{eqnarray*}}
\newcommand{\nn}{\nonumber}
\newcommand{\ba}{\begin{array}}
\newcommand{\ea}{\end{array}}

\DeclareMathOperator{\Tr}{Tr}
\DeclareMathOperator{\Det}{Det}
\DeclareMathOperator{\diag}{diag}
\newcommand{\Phib}{\bar{\Phi}}
\newcommand{\e}{\epsilon}
\newcommand{\B}{{\cal B}}
\newcommand{\F}{{\cal F}}

\newcommand{\D}{{\mathcal D}}

\title{Exact Results 
in
Discretized Gauge Theories}
\author{{}\\So Matsuura${}^1$\footnote{e-mail: s.matsu@phys-h.keio.ac.jp}, 
Tatsuhiro Misumi${}^1$\footnote{e-mail: misumi@phys-h.keio.ac.jp}
and 
Kazutoshi Ohta${}^2$\footnote{e-mail: kohta@law.meijigakuin.ac.jp}
\\[2em]
{\em ${}^1$ Department of Physics in Hiyoshi Campus, Keio University, } \\
{\em 4-1-1 Hiyoshi, Yokohama, 223-8521, Japan} \\
{\em ${}^2$ Institute of Physics, Meiji Gakuin University, Yokohama 244-8539, Japan}}
\date{\today}							

\begin{document}
\maketitle

\begin{center}
{\bf Abstract}
\end{center}

We apply the localization technique to  topologically twisted ${\cal N}=(2,2)$ supersymmetric gauge theory on a
discretized Riemann surface 
(the generalized Sugino model). 
We exactly evaluate the partition function and the vacuum expectation value (vev) 
of a specific $Q$-closed operator.
We show that both the partition function and the vev of the operator
depend only on the Euler characteristic and the area of the discretized Riemann surface 
and are independent of the detail of the discretization. 
This localization technique may not only simplify 
numerical analysis of the supersymmetric lattice models 
but also connect the well-defined equivariant localization 
to the empirical supersymmetric localization.


\newpage

\section{Introduction}

It is known that some of field theories are integrable and we can perform an infinitely dimensional 
path integral completely. In particular, we can exactly obtain a partition function or a part of vacuum 
expectation values (vevs) in two-dimensional Yang-Mills (YM) theories
\cite{Witten:1992xu,Blau:1993hj,Blau:1995rs}
 and three-dimensional
Chern-Simons theories
\cite{Beasley:2005vf,Kapustin:2009kz}, and (extended) supersymmetric YM theories in various dimensions
\cite{Witten:1988ze, Witten:1990bs,Pestun:2007rz}.
A key to understand the integrability of these field theories is the {\it localization} \cite{DH}
(see for review \cite{Karki:1993bw}).
If the localization works in the field theory, the infinitely dimensional path integral reduces
to finite dimensional integrals or discrete sums.
So we can obtain the exact results in this sense.

To validate the localization, 
we need implement a kind of ``supersymmetry'' to the system. 
A typical example of the (non-Abelian) localization appears in
two-dimensional $U(N)$ pure YM theory on an arbitrary 
Riemann surface $\Sigma_h$ with genus $h$ \cite{Witten:1992xu}. 
By introducing an auxiliary scalar field $\Phi$, we can write the partition function as 
\be
Z_{\text{2dYM}} = \int \D \Phi \D A_\mu \, e^{\int_{\Sigma_h} d^2 x \sqrt{g} \Tr [i\Phi F - \frac{g_{\rm YM}^2}{2}\Phi^2]}\,,
\label{2d YM partition function 1}
\ee
where $F$ is the Poincar\'e dual of the field strength, $F=\frac{1}{2}\e^{\mu\nu}F_{\mu\nu}$, 
and $g_{\rm YM}$ is the gauge coupling constant. 
We can obtain the ordinary YM action
$\frac{1}{2g_{\rm YM}^2}\int d^2 x \sqrt{g} \Tr F^2$ by integrating out $\Phi$. 
Here we can introduce fermions (gaugino fields)  $\lambda_\mu$ ($\mu=1,2$)
without changing the value of the partition function (\ref{2d YM partition function 1}),  
\be
Z_{\text{2dYM}} = \int \D \Phi \D A_\mu \D \lambda_\mu \, e^{\int_{\Sigma_h} d^2 x \sqrt{g} \Tr [i\Phi F - \frac{g_{\rm YM}^2}{2}\Phi^2+\lambda_1 \lambda_2]}\,.
\label{2d YM partition function 2}
\ee
We see that the exponent of the integrand of (\ref{2d YM partition function 2}) 
is invariant under the ``supersymmetry'' with a supercharge $Q$
\be
\begin{array}{lcl}
Q \Phi =0, &&\\
Q A_\mu = \lambda_\mu, && Q \lambda_\mu = i\D_\mu \Phi\,.
\end{array}
\label{cont susy1}
\ee
Furthermore, we can regard this symmetry as a part of the supersymmetry of (topologically twisted) 
two-dimensional ${\cal N}=(2,2)$ supersymmetric YM theory if the space-time is flat.
The supersymmetric YM theory also includes extra fields $\Phib$, $Y$, $\eta$ and $\chi$, 
which are transformed as
\be
\begin{array}{lcl}
Q \Phib =\eta, && Q \eta = [\Phi,\Phib],\\
Q Y = [\Phi,\chi], && Q \chi = Y,
\end{array}
\label{cont susy2}
\ee
under the same supercharge.

Using this supercharge, 
the action of the supersymmetric YM theory is written in a $Q$-exact form,
\be
S_{\rm SYM} = \frac{1}{2g_0^2}Q \Xi\,,
\ee
where $\Xi$ is a suitable gauge invariant function of the fields and $g_0$ 
is the coupling constant of the supersymmetric YM.
Using this $Q$-exact action, we can embed the partition function (\ref{2d YM partition function 2})
into the supersymmetric YM theory as a vev of the $Q$-closed ($Q$-invariant) operator
\be
\begin{split}
Z_{\text{2dYM}} &= \int \D \Phi \D A_\mu \D \lambda_\mu \D \Phib \D Y
\D \eta \D\chi \, e^{\int d^2 x \sqrt{g} \Tr [i\Phi F - \frac{g_{\rm YM}^2}{2}\Phi^2+\lambda_1 \lambda_2]+\frac{1}{2g_0^2}Q\Xi}\\
&= \left\langle
e^{\int d^2 x \sqrt{g} \Tr [i\Phi F - \frac{g_{\rm YM}^2}{2}\Phi^2+\lambda_1 \lambda_2]}
\right\rangle_{\rm SYM}\,,
\end{split}
\label{vev in SYM}
\ee
without changing the original value.
In fact, as a consequence of the $Q$-exactness, 
the partition function of two-dimensional YM or the vev in supersymmetric YM (\ref{vev in SYM})
is independent of the coupling constant $g_{0}$ of supersymmetric YM theory.
This means that we can evaluate (\ref{vev in SYM}) exactly in the WKB (1-loop) approximation with respect to the coupling $g_0$ around the fixed points. This is the localization mechanism.
Finally, we get the integral formula of two-dimensional YM
\be
Z_{\text{2dYM}} = \sum_{\vec{k} \in {\mathbb Z}^N}
\int \prod_{i=1}^N \frac{d\phi_i}{2\pi}\prod_{i<j}(\phi_i-\phi_j)^\chi 
e^{\sum_{i=1}^N 2\pi i \phi_i k_i-\frac{g_{\rm YM}^2 {\cal A}}{2}\phi_i^2}\,,
\label{Migdal partition function}
\ee
where $\phi_i$'s are eigenvalues of the adjoint scalar field $\Phi$ and $\chi$ is 
the Euler characteristic of the
Riemann surface $\Sigma_h$ with an area ${\cal A}$.
After taking the summation over the fluxes $k_i$, we obtain Migdal's famous result of two-dimensional YM theory partition function \cite{Migdal:1975zg}.

Our main focus of this paper is a question: 
Does the integrability (localization) of the lower dimensional continuum field theory, 
explained above, still work on a discrete space-time (lattice) or not?
While it is not straightforward to define the whole supersymmetric theory on the lattice because of 
broken translational invariance, 
it is possible to keep a scalar part of extended supersymmetry exact on the lattice, 
which is unaffected by the breaking of translational invariance \cite{Kaplan:2002wv,Cohen:2003xe,Cohen:2003qw,Kaplan:2005ta,Sugino:2003yb,Sugino:2004qd,Sugino:2004uv,Sugino:2006uf,Sugino:2008yp}.
On the other hand, as we have seen in the above, one scalar supercharge enables us to
construct $Q$-exact action and $Q$-closed operators, leading to activation of the localization procedure.
It is thus quite natural to expect that the localization works 
in the lattice supersymmetric YM theory with $Q$-exact action 
\cite{Sugino:2003yb,Sugino:2004qd,Sugino:2004uv,Sugino:2006uf,Sugino:2008yp}, 
and we can obtain some exact results even in the supersymmetric lattice gauge theory.
The localization is earlier applied to the supersymmetric lattice quantum mechanics
in order to calculate the Witten index \cite{Giedt:2004qs}
from the point of view of the Nicolai map on the lattice \cite{Sakai:1983dg,Kikukawa:2002as},
and 
an application of the localization to
the supersymmetric (topological) lattice gauge theory
is first considered in \cite{Ohta:2006qz}.

In this paper, we adopt a generalized version of 
${\cal N}=(2,2)$ supersymmetric lattice gauge theory (Sugino model) 
in which the theory is defined on a discretized Riemann surface \cite{Matsuura:2014kha},
since it is compatible with topologically twisted two-dimensional YM theory where the localization works.
We apply the localization technique to the topologically-twisted theory, and exactly evaluate vacuum expectation values (vev) of some $Q$-closed operators and the partition function itself. 
In particular, we calculate the vev of a physical operator within the theory,
which is a supersymmetrically deformed Kazakov-Migdal (KM) model 
\cite{Kazakov:1992ym, Kogan:1992np}.
We show that the results only depend on Euler characteristic and 
area of the discretized Riemann surface, and are independent of discretization patterns.
Our results are consistent with those for the continuum topologically-twisted ${\cal N}=(2,2)$ supersymmetric gauge theory, which means that the path integral on the lattice partly describes physics in the continuum limit
without lattice artifacts \cite{Witten:1988ze, Witten:1990bs, Witten:1992xu}.

The organization of this paper is as follows:
In the subsequent section, we discuss the localization of a simple unitary matrix model,
which is a famous Harish-Chandra-Itzykson-Zuber (HCIZ) Integral \cite{Harish-Chandra,Itzykson:1979fi},
in prior to considering the Sugino model.
The localization in the HCIZ integral is useful to discuss the localization 
in the supersymmetric lattice gauge theory,
since the lattice gauge theory is essentially multi unitary matrix model.
We first give a review of the Duistermaat-Heckman localization formula 
\cite{DH, Karki:1993bw} for the integral over the unitary group with a suitable Haar measure. 
In section 3,
we consider a direct application of the HCIZ integral to the KM model. 
In section 4, we combine the knowledge of the localization in the unitary matrix models and
apply the localization method to the generalized Sugino model \cite{Sugino:2003yb,Sugino:2004qd,Sugino:2004uv,Sugino:2006uf,Sugino:2008yp}, which is defined on 
the general discretizations of the Riemann surface \cite{Matsuura:2014kha}.
We make use of a supercharge (BRST charge) for the generalized Sugino model, which is
a discretization of the topologically twisted two-dimensional supersymmetric YM theory,
to evaluate the partition function.
We also find that the action of the supersymmetric KM model, 
which is invariant under the BRST supersymmetry, 
surprisingly works as a physical observable in the generalized Sugino model.
The last section is devoted to conclusion and discussion.

\section{Harish-Chandra-Itzykson-Zuber Integral}

\subsection{Equivariant cohomology on coadjoint orbits}

To understand the localization in the lattice gauge theory,
we begin with a simple example of an integrable unitary matrix model.
The localization is originally considered to evaluate a sort of a thermodynamical (classical)
partition function, which is defined by an integral over a phase space
with a symplectic structure. It is known as the Duistermaat-Heckman (DH) localization formula 
\cite{DH,Karki:1993bw}.
We here give a derivation of the localization formula for a specific unitary matrix model,
which is called the Harish-Chandra-Itzykson-Zuber (HCIZ) integral \cite{Harish-Chandra,Itzykson:1979fi}.
We basically follow a review in \cite{Szabo:1996md,Szabo:2000fs}, but some original aspects are added
to clarify important mathematical structures
and connect them with later applications to lattice gauge theory.

Let us now think the following thermodynamical partition function (HCIZ integral)
over a phase space of the unitary group:
\be
Z_{\rm HCIZ} = \int \D U \, e^{-\beta H}\,,
\label{HCIZ integral}
\ee
where
\be
H = \Tr  AUBU^\dag\,,
\label{HCIZ Hamiltonian}
\ee
is regarded as a Hamiltonian written in terms of an $N\times N$ unitary matrix $U$
and Hermitian matrices $A$ and $B$.
The integral of the partition function is defined on a  Haar measure
$\D U$ of the unitary group $U(N)$.
%
%
%
We can generally assume that
the matrices $A$ and $B$ are diagonal; 
$A=\diag(a_1,a_2,\ldots,a_N)$ 
and $B=\diag(b_1,b_2,\ldots,b_N)$,
since the Haar measure is invariant under left and right action onto $U$.

The phase space of the Hamiltonian (\ref{HCIZ Hamiltonian})
is given by
the coadjoint action orbit ${\mathcal O}_B=\{X_B=U B U^\dag|U \in U(N) \}$.
$X_B$ is a ``good'' coordinate on the phase space.
The coadjoint orbit is homeomorphic to the homogeneous coset space of
 $U(N)$ by a maximal torus; ${\mathcal M}=U(N)/U(1)^N$,
 since the matrix $B$ is now diagonal.
${\mathcal M}$ is also called a {\it flag manifold} in the mathematical literature.

${\mathcal M}$ has even dimensions $N(N-1)$ and it is known that ${\mathcal M}$ possesses  a symplectic structure and we can construct a symplectic 2-form on ${\mathcal M}$,
which plays essential role in the localization.

We next consider the equivariant cohomology on ${\mathcal M}$ associated with the HCIZ integral to proceed the localization method.
Let us first consider the left and right invariant 1-forms on ${\mathcal M}$
\be
\theta_L = -iU^\dag dU, \qquad \theta_R=-idUU^\dag\,,
\ee
which are called the Maurer-Cartan (MC) 1-forms.
$\theta_L$ and $\theta_R$ are Hermitian and related with each other by
\be
\theta_L = U^\dag\theta_R U\,.
\ee
We can check that $\theta_L$ and $ \theta_R$ satisfy the Maurer-Cartan equation
\bea
d\theta_L &=& -i \theta_L \wedge \theta_L,\\
d\theta_R &=& +i\theta_R \wedge \theta_R\,.
\label{MC eq R}
\eea
We can see the exterior derivative on the coordinate $X_B$ becomes 
\be
d X_B = i[\theta_R,X_B]\,.
\ee
Thus we find that
the exterior derivative of the Hamiltonian $H=\Tr AX_B$ is proportional to $\theta_R$
\be
dH = i\Tr
[X_B,A]\theta_R\,.
\label{dH}
\ee

Using the right invariant MC 1-form,
we can define the symplectic 2-form on ${\mathcal M}$,
which is called Kirillov-Kostant-Souriau
symplectic 2-form \cite{nla.cat-vn26429,Kostant,souriau1997structure}, at a point $X_B$
\be
\omega(X_B)  = \Tr\left(X_B \theta_R \wedge \theta_R \right)\,.
\ee
Then we find that $\omega(X_B)$ is closed, namely $d\omega(X_B)=0$\,.

The Hamiltonian and symplectic 2-form on the phase space
define the Hamiltonian vector field $V$ by the equation
\be
dH = \iota_V \omega\,,
\label{definition of the vector field}
\ee
where $\iota_V$ stands for the interior product with respect to $V$.
Comparing (\ref{dH}) with
\bea
\iota_V \omega &=& \Tr\left(
X_B (\iota_V \theta_R)\theta_R
-X_B \theta_R (\iota_V \theta_R)
\right)\nn\\
&=&
\Tr\left(
[X_B, \iota_V \theta_R]\theta_R
\right)\,,
\eea
we find that $\iota_V \theta_R=iA$.
The fixed points of the Hamiltonian vector flow $V=0$ given by an equation
$dH=0$, that means $[X_B,A]=0$.
In terms of $U$, the fixed points are given by a permutation group
$\Gamma_\sigma$, labelled by a permutation $\sigma$, in the group $U(N)$.
In the next subsection, we will show that the fixed points of the
Hamiltonian vector flow are of significance in the integral, 
and the integral (\ref{HCIZ integral}) localizes at these fixed points.

The equivariant differential operator is defined by
\be
d_V \equiv d + \iota_V\,,
\ee
which constructs the equivariant cohomology on ${\mathcal M}$.
In particular,
we find that $H-\omega$ is an element of the equivariant cohomology class, since
we see immediately
$d_V (H-\omega) = 0$ from the definition of the Hamiltonian vector field (\ref{definition of the vector field}).
We also find an algebra for the basic variables
\be
\ba{lcl}
d_V U U^\dag =  i\theta_R,& &d_V \theta_R = i A+ i \theta_R \wedge \theta_R\,,
\ea
\label{equivariant alg}
\ee
where we have used the MC equation (\ref{MC eq R}).

Using the symplectic structure of ${\mathcal M}$ and the equivariant cohomology generated
by $d_V$, we can mathematically develop  the localization theorem with respect to the HCIZ integral.
However our purpose in this paper is to understand it in terms of the localization in the
supersymmetric system. So we introduce the ``supersymmetry'' to the HCIZ integral
and relate it with the equivariant cohomology in the next subsection.

\subsection{Supersymmetry}

It is known that 1-forms in the differential geometry is naturally identified with fermionic variables
(Grassmann numbers).
We here identify the MC 1-form $\theta_R$ with a Grassmann valued
(fermionic) variable $\lambda_R$.
Note that the symplectic 2-form becomes $\omega(X_B)=-\frac{1}{2}\Tr\lambda_R[X_B,\lambda_R]$
under this identification.
If we also identify $d_V$ with a supercharge $Q$,
the algebra (\ref{equivariant alg}) gives a relation among bosonic and fermionic variables,
that is, 
the BRST symmetry (supersymmetry) 
\be
\ba{lcl}
Q U =  i\lambda_R U,& &Q \lambda_R = i A+ i \lambda_R \lambda_R\,.
\label{original BRST}
\ea
\ee
Of course, $Q(H-\omega)=0$ is satisfied.
This symmetry plays a crucial role in the localization.

Let us now go back to the HCIZ integral (\ref{HCIZ integral}).
Incorporating $\lambda_R$, $\lambda_L$ and $\omega(X_B)$, the HCIZ integral is written by
\be
Z_{\rm HCIZ} = \frac{1}{\beta^{N(N-1)/2}\Delta(b)}\int \D U \D \lambda_L \, e^{-\beta (H - \omega)},
\label{Zomega}
\ee
where $\Delta(b) \equiv \prod_{i<j}(b_i-b_j)$ is a Vandermonde determinant of 
the eigenvalues of $B$.
We also have removed Cartan parts of the bosonic and fermionic
integral variables because of the quotient space of
the phase space.
The normalization factor
in (\ref{Zomega}) is determined by 
the integral of $\omega$ over the fermions as 
\bea
\int \D \lambda_L \, e^{\beta\omega} &=& \int \D \lambda_L \, e^{-\frac{\beta}{2}\Tr \lambda_R [X_B, \lambda_R]}\nn\\
&=&\int \D \lambda_L \, e^{-\frac{\beta}{2}\Tr  \lambda_L [B, \lambda_L]}\nn\\
&=&\beta^{N(N-1)/2}\Delta(b)\,,
\eea
where we have fixed the integral measure by the fermionic variable $\lambda_L$
instead of $\lambda_R$,
in order to avoid signatures depending on the Weyl group (permutations) in $U$.


Noting that $H -\omega$ is $Q$-closed, the integral can be deformed by a $Q$-exact term
\be
Z_t = \frac{1}{\beta^{N(N-1)/2}\Delta(b)}\int \D U \D \lambda_L \, e^{-\beta( H -\omega) - t Q \Xi}\,,
\label{Zomega2}
\ee
without changing the value of the integral,
since the deformed integral is independent of the parameter $t$: 
\be
\frac{\delta Z_t}{\delta t} = 
-\frac{1}{\beta^{N(N-1)/2}\Delta(b)}\int \D U \D \lambda_L \, Q( \Xi e^{-\beta (H -\omega) - t Q \Xi})=0\,,
\ee
under the $Q$-invariant measure of the integral if $\Xi=0$ at integral boundaries.
Thus we can evaluate the integral exactly by using the saddle point (fixed point) approximation with respect to the $Q$-exact term in the limit of $t \to \infty$.

Here we should note that $H-\omega$ itself is written as a $Q$-exact form
\be
\begin{split}
-iQ\Tr  X_B\lambda_R
&=\Tr\left(
[\lambda_R,X_B]\lambda_R
+ AX_B+X_B\lambda_R  \lambda_R
\right)\\
&=
 H-\omega\,.
\end{split}
\ee
However this does not immediately mean that the integral (\ref{HCIZ integral})
is independent of the parameter (inverse temperature) $\beta$,
since $\Xi'=\Tr  X_B\lambda_R$ takes a non-zero value at the
boundary of the integration domains.
So we should find another ``good'' $Q$-exact term in order to 
utilize the saddle point approximation to the HCIZ integral.

According to the general argument in the localization theorem \cite{Karki:1993bw},
the extra $Q$-exact term should provide the same equation of motion as
the original Hamiltonian $H$. This copy of the Hamiltonian system is
called the bi-Hamiltonian structure.

In the following arguments to construct the bi-Hamiltonian structure, it is useful to define a
new fermionic variable  $\Lambda_B\equiv i[\lambda_R,X_B]$ associated with the coordinate $X_B$
on ${\mathcal M}$.
The supersymmetry transformations among these variables become
\be
\ba{lcl}
Q X_B =  \Lambda_B,& &Q \Lambda_B = -[A,X_B]\,.
\ea
\ee
If we choose now $\Xi$ as follows
\be
\Xi = -\frac{1}{2}\Tr \Lambda_B (Q\Lambda_B)\,,
\ee
where we have defined, then 
we obtain
\be
Q\Xi =  K- \Omega\,,
\ee
where $K=-\frac{1}{2} \Tr (Q\Lambda_B)^2 = -\frac{1}{2}\Tr[A,X_B]^2$ and $\Omega=\frac{1}{2}\Tr \Lambda_B[A,\Lambda_B]$.
We see that $K$ and $\Omega$ possess the same Hamiltonian structure as the original one,
that is, $(H,\omega)$ and $(K,\Omega)$ provide the bi-Hamiltonian structure.

Using the $t$-independence of the integral (\ref{Zomega2}),
we can take the limit $t\to \infty$ without changing the value of the integral,
and then the saddle point approximation
with $(K,\Omega)$ becomes exact.
Each solution of the saddle point equation $[A,X_B]=0$ is labelled by
the permutation and $U=\Gamma_\sigma$ as we mentioned.
If we denote $U=e^{\frac{i}{\sqrt{t}}Z}\Gamma_\sigma$ by using
a fluctuation $Z$, which is a Hermitian matrix, around the saddle point,
$X_B$ is expanded as
\be
X_B \simeq B_\sigma + \frac{i}{\sqrt{t}}[Z,B_\sigma]\,,
\ee
where $B_\sigma \equiv \Gamma_\sigma B \Gamma_\sigma^\dag
= \diag(b_{\sigma(1)}, b_{\sigma(2)},\ldots,b_{\sigma(N)})$, and
\be
\Lambda_B \simeq 0 + \frac{i}{\sqrt{t}}\Gamma_\sigma [\tilde{\lambda}_L,B]\Gamma_\sigma^\dag\,,
\ee
where $\tilde{\lambda}_L$ is a fluctuation of the integral variable $\lambda_L$.
Substituting these expansion into $K$ and $\Omega$, 
we get
\bea
t K &=& \frac{1}{2}\Tr [A,[B_\sigma,Z]]^2+{\cal O}(1/\sqrt{t}),\\
t \Omega&=&-\frac{1}{2}\Tr [B,\tilde{\lambda}_L][A_\sigma,[B,\tilde{\lambda}_L]] +{\cal O}(1/\sqrt{t})\,,
\eea
where
$A_\sigma \equiv \Gamma_\sigma^\dag A \Gamma_\sigma
= \diag(a_{\sigma(1)}, a_{\sigma(2)},\ldots,a_{\sigma(N)})$.

Performing the Gaussian integrals over $Z$ and $\tilde{\lambda}_L$
in the $t\to \infty$ limit, we obtain the following exact integral result
as a summation over the saddle points (permutations)
\bea
Z
&=& \left(\frac{2\pi}{\beta}\right)^{N(N-1)/2}
\frac{1}{\Delta(b)}\sum_{\sigma} 
\frac{\Delta(a_\sigma)\Delta(b)^2}{|\Delta(a)|^2|\Delta(b_\sigma)|^2}
e^{-\beta \sum_i a_i b_{\sigma(i)}}\\
&=& \left(\frac{2\pi}{\beta}\right)^{N(N-1)/2}\sum_{\sigma} 
(-1)^{|\sigma|}
\frac{e^{-\beta \sum_i a_i b_{\sigma(i)}}}{\Delta(a)\Delta(b)}\\
&=& \left(\frac{2\pi}{\beta}\right)^{N(N-1)/2}
\frac{\det_{i,j} e^{-\beta a_i b_j}}{\Delta(a)\Delta(b)}\,,
\eea
where $\Delta(a)=\prod_{i<j}(a_i-a_j)$ is a Vandermonde determinant for $A$,
and $\Delta(a_\sigma)$ and $\Delta(b_\sigma)$ are those for the permuted eigenvalues.
We have also used the fact that
$\Delta(a_\sigma)/\Delta(a)=(-1)^{|\sigma|}$,
which gives the signature of the permutation.
This agrees with the known result \cite{Harish-Chandra,Itzykson:1979fi}.

Before closing this section, we would like to point out that we can modify the
BRST transformation (\ref{original BRST}) by adding a term which commute with $X_B$
without changing the above localization argument and the $Q$-exact ``action'' $K-\Omega$.
For example, up to a linear term, we can modify (\ref{original BRST}) as
\be
\ba{lcl}
Q U =  i\lambda_R U,& &Q \lambda_R = i (A -q X_B+ \lambda_R \lambda_R),
\ea
\ee
where $q$ is a constant parameter.
This redundant symmetry in the BRST transformation
will be important in the argument of the supersymmetric lattice gauge theory.

\section{Kazakov-Migdal Model}

In Ref.~\cite{Kazakov:1992ym}, 
Kazakov and Migdal proposed an intriguing lattice gauge (multi matrix) model 
with the action, 
\be
S_{\rm KM} = -\sum_{\langle xy\rangle} \Tr \Phi_x U_{xy} \Phi_y U_{xy}^\dag
+\sum_x \Tr V(\Phi_x)\,,
\label{KM model}
\ee
where $x$ and $y$ represent the lattice points,
and $\langle xy \rangle$ denotes the nearest neighbor links between $x$ and $y$.
The unitary matrices $U_{xy}$ are defined on each link $\langle xy \rangle$
and the Hermitian matrix $\Phi_x$ is on each site $x$.
The potential is mostly chosen to be a quadratic one $V(\Phi_x)=\frac{m}{2}\Phi_x^2$.
This model is called the Kazakov-Migdal (KM) model and is originally constructed 
in order to induce the YM theory in any dimensions.

The action (\ref{KM model}) has almost the same form as the previous HCIZ Hamiltonian,
except for the potential term and that $\Phi_x$'s are not constant matrices but now integral variables
in the path integral. The partition function of the KM model is
\be
Z_{\rm KM} = \int \prod_x \D \Phi_x \prod_{\langle xy\rangle} \D U_{xy}
\, e^{-S_{\rm KM}}\,.
\ee
Integrating out all adjoint scalar fields $\Phi_x$, we obtain an effective action
which mimics the Yang-Mills theory in the continuum limit.
On the other hand, if we integrate the unitary link variables
by using the HCIZ integral, we obtain a multiple integral over the eigenvalues 
$(\phi_{x,1},\phi_{x,2},\ldots,\phi_{x,N})$ of $\Phi_x$
\be
Z_{\rm KM} = \int \prod_x \prod_{i=1}^N d \phi_{x,i} \, e^{-V(\phi_{x,i})}
\Delta(\phi_x)^2
\prod_{\langle xy\rangle} I_{xy}\,,
\label{KM result}
\ee
where $\Delta(\phi_x)^2=\prod_{i<j}(\phi_{x,i}-\phi_{x,j})^2$
comes from the measure of $\Phi_x$ in the diagonal gauge
as well as the Hermitian matrix model, and
$I_{xy}$ is the result of the HCIZ integral\footnote{We ignore irrelevant overall constants in the
partition function.}
\be
I_{xy} = \frac{\det_{i,j} e^{\phi_{x,i}\phi_{y,j}}}{\Delta(\phi_x)\Delta(\phi_y)}\,.
\ee

As we discussed in the previous section, the integrability of the HCIZ integral
is essentially caused by the localization with the supersymmetry.
Since the KM model has almost the same structure as the HCIZ integral,
we can introduce fermionic variables $\lambda_{xy}$ with 
the following transformation under the action of the supercharge, 
\be
\ba{lcl}
Q U_{xy} =  i\lambda_{xy} U_{xy},& &Q \lambda_{xy} = i \Phi_x + i \lambda_{xy} \lambda_{xy},\\
Q \Phi_x =  0.& &\\
\ea
\ee
Note that $\lambda_{xy}$ lives on the site $x$ of the link $\langle xy \rangle$. 
Unfortunately, the action (\ref{KM model}) itself is not invariant under the above symmetry,
namely $QS_{\rm KM}\neq 0$, so we ``supersymmetrize'' the action by adding
a fermionic term corresponding to the symplectic 2-form on the coadjoint orbit
\be
S_{\rm sKM} = -\sum_{\langle xy\rangle} 
\Tr \left\{
\Phi_x U_{xy} \Phi_y U_{xy}^\dag
-\frac{1}{2} \lambda_{xy} [U_{xy} \Phi_y U_{xy}^\dag,\lambda_{xy}]
\right\}
+\sum_x \Tr V(\Phi_x)\,.
\label{sKM action}
\ee
We can easily check $QS_{\rm sKM}=0$ and refer to this action as the supersymmetric Kazakov-Migdal (sKM) model
in the following.

We first integrate over $U_{xy}$ and $\lambda_{xy}$ of the partition function of the
sKM model
\begin{align}
Z_{\rm sKM} =& \int \prod_x \D \Phi_x \D \lambda_{xy} \prod_{\langle xy\rangle} \D U_{xy}
\, e^{-S_{\rm sKM}} \nn \\
=& \int \prod_x \prod_{i=1}^N d \phi_{x,i} \, e^{-\sum_{x,i} V(\phi_{x,i})}
\Delta(\phi_x)^2
\prod_{\langle xy\rangle} \Delta(\phi_y) I_{xy},
\label{sKM result}
\end{align}
which is slightly different from (\ref{KM result}) 
by the number of the Vandermonde determinants.
Repeating
the localization argument,
we can construct a $Q$-exact action
for the multi unitary matrix model (sKM)
\be
\begin{split}
Q\Xi &= -\frac{1}{2}Q\sum_{x}\Tr [\lambda_{xy},U_{xy} \Phi_y U_{xy}^\dag]
[\Phi_x,U_{xy} \Phi_y U_{xy}]\\
&=-\frac{1}{2}
\sum_{x}\Tr \left\{
[\Phi_x,U_{xy} \Phi_y U_{xy}]^2
-\frac{1}{2}[\lambda_{xy},U_{xy} \Phi_y U_{xy}^\dag]
[\Phi_x,[\lambda_{xy},U_{xy} \Phi_y U_{xy}^\dag]]
\right\}\,.
\end{split}
\label{Q-exact action for sKM}
\ee
Then we can deform the partition function (\ref{sKM result})
by the $Q$-exact action (\ref{Q-exact action for sKM}) without changing 
the value of the partition function as
\be
\begin{split}
Z_{\rm sKM} &= \int \prod_x \D \Phi_x \D \lambda_{xy} \prod_{\langle xy\rangle} \D U_{xy}
\, e^{-S_{\rm sKM}- t Q\Xi}\\
&= \left\langle
e^{-S_{\rm sKM}}
\right\rangle\,.
\end{split}
\ee
Thus we can regarded
the partition function of the sKM model
as the vev of the $Q$-closed operator $e^{-S_{\rm sKM}}$
in the theory with the action $Q\Xi$.

This seems to be a counterpart of the localization argument in the continuum field theory
as explained in Introduction.
However, in the continuum limit,
the discretized action (\ref{Q-exact action for sKM}) does not coincide with
the (topologically twisted) ${\cal N}=(2,2)$ supersymmetric YM action.
Indeed the action (\ref{Q-exact action for sKM}) may not reflect the
symmetry of the two-dimensional YM theory. (Recall that the original 
KM model defines the discretized theory in {\it any} dimensions.)
In order to conform to the two-dimensional YM theory,
we need to introduce extra fields as well as in the supersymmetric continuum
YM theory.
We discuss a different type of the discretized action from the above
in the next section.

\section{${\mathcal N}=(2,2)$ Supersymmetric Lattice Gauge Theory}

\subsection{Generalized Sugino model}

So far, we have considered exact solvable unitary matrix models via the localization.
In this section, we reverse the above arguments by introducing a 
two-dimensional supersymmetric lattice
model on a discretization of Riemann surfaces \cite{Matsuura:2014kha}.
As we will see below, $S_{\rm sKM}$ works as a $Q$-closed physical observable 
in this supersymmetric lattice theory.

\begin{figure}[ht]
\begin{center}
\includegraphics[scale=0.8]{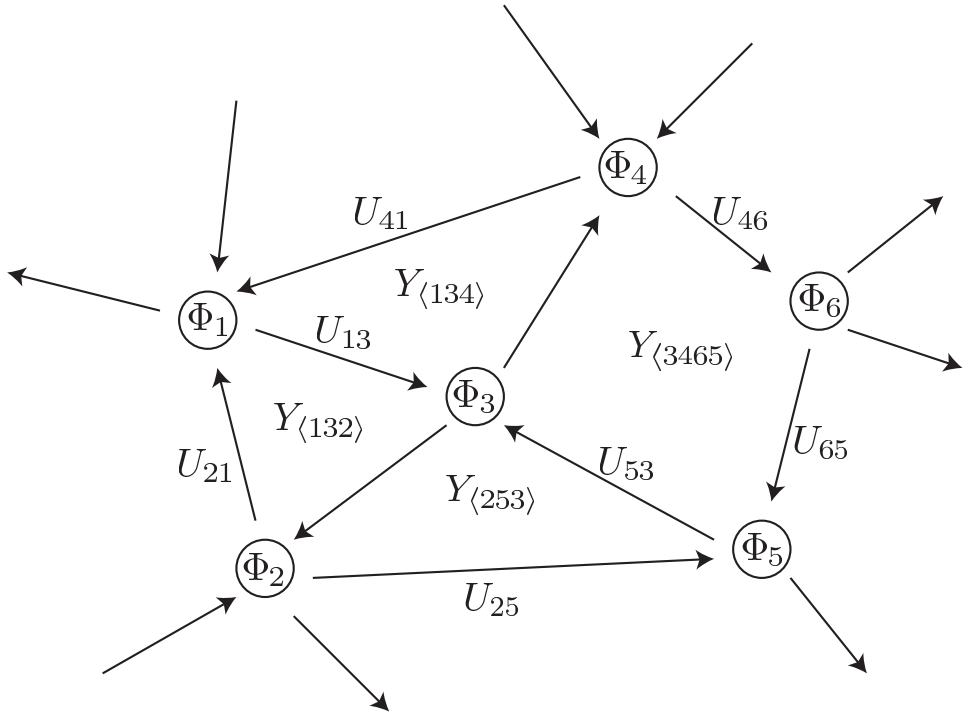}
\end{center}
\caption{The structure of the generic lattice. We here show only the bosonic fields
associated with each site, link and face.}
\label{lattice diagram}
\end{figure}

Following \cite{Matsuura:2014kha}, we first discretize the Riemann surface
by gluing together two-dimensional polygons with points (sites) and edge lines (links).
We denote a set of sites, links and faces by $S$, $L$ and $F$, respectively.
We assume that each link is oriented. 
Once we define such a generic lattices (discretized space-time),
we can construct the ${\cal N}=(2,2)$ supersymmetric discretized gauge theory on it 
by assigning scalar fields $\Phi_x$ on the sites, unitary matrices $U_{xy}$ 
on the links $\langle xy \rangle$,
auxiliary fields $Y_f$ on the faces, fermions $\lambda_{xy}$, $\eta_x$ on the sites,
and fermions $\chi_f$ on the faces. (See Fig.~\ref{lattice diagram}.)

We now introduce the BRST (supersymmetry) transformation for these variables by
\be
\begin{array}{ll}
Q_s\Phi_x =0, &  \\
Q_s \Phib_x = \eta_x, & Q_s\eta_x=[\Phi_x,\Phib_x]\\
Q_s U_{xy} = i\lambda_{xy} U_{xy},
& Q_s \lambda_{xy}=i(U_{xy}\Phi_yU_{xy}^\dag-\Phi_x+ \lambda_{xy}\lambda_{xy}),\\
Q_s Y_f =[\Phi_f,\chi_f],
 & Q_s \chi_f = Y_f, 
\end{array}
\label{Sugino BRST transformation}
\ee
which are the lattice analog of (\ref{cont susy1}) and (\ref{cont susy2}). 
Here we denote the BRST charge by $Q_s$ to distinguish from the previous one.
On all variables, the transformation satisfies
$Q_s^2\cdot =i\delta_\Phi \cdot$ where $\delta_\Phi$ denotes an infinitesimal
gauge transformation with the parameter $\Phi$.
For later convenience, we define fermions on the links by $\Lambda_{xy}\equiv \lambda_{xy} U_{xy}$.
Then the third line of the BRST transformation reduces to
\be
Q_s U_{xy} = i\Lambda_{xy}, \quad
 Q_s \Lambda_{xy}=i(U_{xy}\Phi_y- \Phi_xU_{xy})\,.
\ee

Using the above BRST transformation, 
the action can be written in a $Q_s$-exact form
\be
S =\frac{1}{2g_0^2}Q_s\left[\sum_{x\in S} \alpha_x \Xi_x
+\sum_{\langle xy \rangle\in L} \alpha_{\langle xy \rangle} \Xi_{\langle xy \rangle}
+\sum_{f\in F} \alpha_f \Xi_f\right]\,,
\label{Q-exact Sugino action}
\ee
with
\begin{align}
\Xi_x &\equiv \Tr \biggl\{
\frac{1}{4}\eta_x [\Phi_x,\Phib_x]
\biggr\},\\
\Xi_{\langle xy \rangle} &\equiv \Tr \biggl\{
-i \Lambda_{xy}(\Phib_y U_{xy}^\dag - U_{xy}^\dag\Phib_x)
\biggr\},\\
\Xi_f &\equiv \Tr \biggl\{
\chi_f(Y_f-i\beta_f \mu(U_f))
\biggr\}\,,
\end{align}
where the coupling constants $\alpha_x$, $\alpha_{\langle xy \rangle}$, $\alpha_f$ and $\beta_f$
should be fixed in order to reproduce a correct continuum limit of the topological field theory 
\cite{Matsuura:2014kha}.
However, surprisingly, the partition function and the vev of some 
physical observables are independent of them 
as we will see.
The theory is constrained on $\mu(U_f)=0$ after integrating out the auxiliary fields,
where $\mu(U_f)$ is a function of a plaquette variable $U_f$
defined by
\be
U_f \equiv \prod_{i=0}^{n} U_{x_i x_{i+1}}\,,\quad (x_{n+1}=x_0)
\ee
where $f=\langle x_0 x_1 \cdots x_n \rangle$
is the face surrounded by the links $\langle x_0 x_1 \rangle, \ldots, \langle x_n x_0 \rangle$.
The function $\mu(U_f)$ is associated with the D-term constraint (moment map)
in the continuum theory.
In the lattice gage theory, 
we can choose $\mu(U_f)$ so that $U_f=1$ is the unique solution of 
the vacuum equation $\mu(U_f)=0$. 
For detail see Ref.~\cite{Matsuura:2014pua}.
%
%
%
%
After acting $Q_s$ in (\ref{Q-exact Sugino action}), we obtain the explicit form of the action 
\be
\begin{split}
S&=
 \frac{1}{2g_0^2}\Tr\Biggr[
\sum_{x\in S} \frac{\alpha_x}{4}[\Phi_x,\Phib_x]^2
+\sum_{\langle xy \rangle \in L} \alpha_{\langle xy \rangle}|U_{xy}\Phi_y-\Phi_x U_{xy}|^2
+\sum_{f\in F} \alpha_f Y_f(Y_f-i\beta_f\mu (U_f))\\
&\qquad\qquad\quad-\sum_{x\in S} \frac{\alpha_x}{4}\eta_x[\Phi_x,\eta_x]\\
&\qquad\qquad\quad +\sum_{\langle xy \rangle \in L} i \alpha_{\langle xy \rangle} \Lambda_{xy}(\eta_x U_{xy}^\dag-U^\dag_{xy}\eta_x
-\Phib_{y}U_{xy}^\dag\Lambda_{xy}U_{xy}^\dag+U_{xy}^\dag\Lambda_{xy}U_{xy}^\dag\Phib_x)\\
&\qquad\qquad\quad +\sum_{f\in F}\alpha_f \left(-\chi_f[\Phi,_f\chi_f]+i\chi_f \beta_f Q_s\mu (U_f)\right)
\Biggr]\,.
\end{split}
\ee

\subsection{Localization and exact results}

To proceed the localization argument, we first show the partition function is independent of the coupling constants;
$g_0$, $\alpha_x$, $\alpha_{\langle xy \rangle}$, $\alpha_f$ and $\beta_f$.
First of all, noting that we can always rescale pairs of  the variables $(\bar{\Phi}_x,\eta_x)$ and
$(Y_f,\chi_f)$ without changing the measure because of the supersymmetry.
This means that the partition function is invariant under the change of the coupling constants
\be
\begin{array}{lcl}
\alpha_x \to c_1^2 \alpha_x, && 
\alpha_{\langle xy \rangle} \to c_1 \alpha_{\langle xy \rangle} ,\\
\alpha_f \to c_2^2  \alpha_f, && \beta_f  \to c_2 \beta_f,
\end{array}
\label{coupling relation}
\ee
with constants $c_1$ and $c_2$.
In addition, we can show that the partition function is completely independent
of the couplings $\alpha_x$ and $\alpha_f$,
since the action constructing from $\Xi_x$ and the first term of $\Xi_f$
is essentially Gaussian and there is no contribution from the moduli boundary.
Combining them, we see that the partition function is independent of all of the
coupling constants. 
The independence of the overall coupling $g_0$ is apparent since
we can always include $g_0$ to the others.

Using the coupling independence, we choose all of coupling
to be $\alpha_x=\alpha_{\langle xy \rangle}=\alpha_f=\beta_f=1$,
except for the overall coupling $g_0$, in the following.
Then
the $Q_s$-exact action can be simply written as
\be
S = \frac{1}{2g_0^2}Q_s\Tr\Big[
\vec{\F}\cdot \overline{Q_s\vec{\F}} -i \sum_{f\in F} \chi_f \mu(U_f)
\Big]\,,
\label{gen action}
\ee
where
we have introduced the sets of bosonic and fermionic fields
$\vec{\B} = (\Phib_x,U_{xy},Y_f)$ and
$\vec{\F} = (\eta_x,\Lambda_{xy},\chi_f)$, 
respectively, and
 `` $\cdot$ '' denotes a suitable inner product with summation over corresponding variables
associated with the lattice structure. 
Thus we can regard the supersymmetric lattice gauge theory as a supersymmetric Gaussian matrix model
with a constraint by the moment maps $\mu(U_f)=0$.\footnote{
Here we should note that $\bar\Phi_x$ is {\it not} the Hermitian conjugate of 
$\Phi_x$ but an independent Hermitian variable. 
Thus the symbol $\overline{\cdots}$ in the expression (\ref{gen action}) do not mean 
to take the Hermitian conjugate but merely exchange $\Phi_x$ and $\bar\Phi_x$. 
}
Moreover, using the coupling independence of $g_0$,
we find that
the partition function and vev of physical observables
are exactly evaluated at the 1-loop level, and
the path integral is localized at the set of the BRST fixed point $Q\vec{\cal F}=0$
and the moment map constraint $\mu(U_f)=0$.

In evaluating the partition function, we first fix the gauge by diagonalizing 
$\Phi_x$ as
\begin{equation}
\Phi_x = \diag(\phi_{x,1},\phi_{x,2},\ldots,\phi_{x,N}).
\label{diag gauge}
\end{equation}
Note that this gauge breaks the gauge group from $\prod_{x\in S}U(N)$ to $\prod_{x\in S}U(1)^N$.
The most nontrivial BRST fixed point condition is that for the link fermions, 
\begin{equation}
U_{xy}\Phi_y - \Phi_x U_{xy} = 0 \quad \text{for }{}^\forall \langle xy \rangle \in L,
\label{fixed point eqs}
\end{equation}
which can be solved by 
\be
U_{xy}=\Gamma_{xy} \in \mathfrak{S}_N,
\ee
where $\mathfrak{S}_N$ is the permutation (Weyl) subgroup in $U(N)$,
since $\Phi_x$ is diagonal.
Thus we find that the diagonal elements of $\Phi_x$ 
between neighbor nodes are related with each other by the permutations
\be
\Phi_{y} = \Gamma_{xy}^\dag \Phi_x \Gamma_{xy}.
\ee
This means that all the eigenvalues of $\Phi_x$ are expressed by permutations 
of a representative eigenvalue at some point.
If we denote the representative eigenvalue by $\phi_i$,
the other eigenvalues are determined by a permutation 
of it, namely
\be
\phi_{x,i} = \phi_{\sigma_x(i)},
\ee
where $\sigma_x(i) \in \mathfrak{S}_N$
and we have assumed that all the sites are connected.

In addition, the moment map constraint $\mu(U_f)=0$ requires 
\be
\left.U_{f}\right|_{U_{xy}=\Gamma_{xy}}=1,
\ee
which is also a consistency condition of the permutations around any face.
So we can choose sets of the possible permutations 
which satisfy the constraint by each face $f$.
Thus the eigenvalue at each point is also determined by the chain of the possible permutations from the
representative point.

In evaluating the partition function in the saddle point approximation, 
we have to compute the 1-loop determinant (Jacobian of the Gaussian integrals) 
around the fixed points, 
which is obtained as the determinant of
the super Hessian matrix (see \cite{Ohta:2012ev,Ohta:2013zna} and Appendix A)
\begin{align}
\text{(1-loop det)}
&=\sqrt{\frac{\Det'\frac{\delta Q_s\vec{\B}}{\delta \vec{\F}}}{\Det'\frac{\delta Q_s\vec{\F}}{\delta \vec{\B}}}}\,,
\label{1-loop det}
\end{align}
where each determinant is taken only over non-zero modes (non-zero eigenvalues).
Here we have to carefully remove the zero modes in the determinant to avoid zeros or divergences.
Evaluating the above 1-loop determinant of our model in the diagonal gage, we find
\be
\begin{split}
\text{(1-loop det)}
&=\sqrt{
\frac{\frac{\delta Q_s \eta_x}{\delta \bar{\Phi}_x}}
{\frac{\delta Q_s \eta_x}{\delta \bar{\Phi}_x}\frac{\delta Q_s U_{xy}}{\delta \Lambda_{xy}}
}}\\
&=\sqrt{
\frac{\prod_{f\in F}\prod_{\sigma_f(i)\neq \sigma_f(j)}(\phi_{\sigma_f(i)}-\phi_{\sigma_f(j)})}
{
\prod_{x\in S}\prod_{\sigma_x(i)\neq \sigma_x(j)}(\phi_{\sigma_x(i)}-\phi_{\sigma_x(j)})
\prod_{\langle xy \rangle \in L}\prod_{\sigma_x(i)\neq \sigma_y(j)}(\phi_{\sigma_x(i)}-\phi_{\sigma_y(j)})
}
},
\end{split}
\ee
where $\phi_{\sigma_f(i)}$ stands for an eigenvalue at an {\it arbitrary} point on the face $f$.

In addition to the above  1-loop determinant, we also need the Vandermonde determinant
at each point
$\prod_{x\in S}\prod_{\sigma_x(i)\neq \sigma_x(j)}(\phi_{\sigma_x(i)}-\phi_{\sigma_x(j)})$,
which appears
in the integration of the gauge fixing ghosts.
Combining the 1-loop determinant with the Vandermonde determinant,
we obtain the partition function as an integration over the representative eigenvalue
and a summation over the possible permutations (fixed points)
\be
\begin{split}
Z &= \sum_{\sigma_{x}:\text{possible permutations}} 
\int \prod_{i=1}^N \frac{d\phi_{i}}{2\pi i} \\
&\qquad \times\sqrt{
\frac{
\prod_{x\in S}\prod_{\sigma_x(i)\neq \sigma_x(j)}(\phi_{\sigma_x(i)}-\phi_{\sigma_x(j)})
\prod_{f\in F}\prod_{\sigma_f(i)\neq \sigma_f(j)}(\phi_{\sigma_f(i)}-\phi_{\sigma_f(j)})}
{\prod_{\langle xy \rangle \in L}\prod_{\sigma_x(i)\neq \sigma_y(j)}(\phi_{\sigma_x(i)}-\phi_{\sigma_y(j)})
}}.
\end{split}
\ee
Using the fact that the difference product of the eigenvalues in the integrand
is invariant under the permutations
and 
the contributions to the measure from each permutation are identical,\footnote{
One might think that some signs (phases) appear in the permutations, but
the whole of integrand should be invariant under the permutations since
the permutation group is a part of the original gauge symmetry $U(N)$.
}
we finally obtain a simple expression of the partition function
\be
Z ={\cal C} \int \prod_{i=1}^N \frac{d\phi_i}{2\pi i} 
\prod_{i<j}(\phi_i-\phi_j)^{n_S - n_L + n_F}, 
\label{result of the partition function}
\ee
where $n_S$, $n_L$ and $n_F$ are the numbers of sites, links and faces, respectively, 
and 
${\cal C}$ is the total number of the possible permutations.
We here would like to emphasize that the original path integral of the lattice gauge theory reduces to
an integral over only $N$ eigenvalues at the representative point,
thanks to the localization.


Here the combination 
$\chi \equiv n_S - n_L + n_F$ is nothing but the Euler characteristic
which depends only on the topology of the two-dimensional surface. 
It is remarkable that the final result of the partition function 
(\ref{result of the partition function}) is the same as the
partition function of (topologically twisted) ${\cal N}=(2,2)$ supersymmetric Yang-Mills theory on the smooth Riemann surface ({\it continuum} space-time)
\cite{Witten:1992xu,Blau:1993hj,Blau:1995rs}.
The integral (\ref{result of the partition function}) of the partition function diverges in general for $\chi\geq 0$.
This fact reflects the existence of the flat direction of the supersymmetric theory.
In order to regularize the divergence from the flat direction, we need to turn on a potential
without spoiling the localization argument. This is done by introducing physical observables (BRST closed operators)
as we will discuss in the next subsection.

Before going to the next subsection, we mention that 
there is an alternative way to take into account the zero-modes at the fixed points
by using a residue integral over eigenvalues of $\Phi_x$'s.
To see this, let us go back to the original expression of the partition 
function (\ref{gen action}) in the diagonal gauge (\ref{diag gauge}). 
Since we can use the formula for the 1-loop determinant (\ref{1-loop det}) 
before localizing the path integral over $\Phi_x$, 
we obtain 
\be
Z = \int \prod_{x\in S} \prod_{i=1}^N \frac{d\phi_{x,i}}{2\pi i}
\frac{\prod_{x\in S}\prod_{i<j}(\phi_{x,i}-\phi_{x,j})\prod_{f\in F}\prod_{i<j}(\phi_{f,i}-\phi_{f,j})}
{\prod_{\langle xy \rangle \in L}\prod_{i\leq j}(\phi_{x,i}-\phi_{y,j})}\,. 
\label{partition function}
\ee
To integrate the diagonal elements of $\Phi_x$, 
we need to choose suitable contours for each $\phi_{x,i}$,
which corresponds to the gauge fixing of the residual $U(1)$'s and moment map constraints
\cite{Ohta:2014ria}.
By choosing the contours and picking up the poles in the integral (\ref{partition function}),
we obtain an integral results as a residue integral.
The poles of the integrand exactly 
correspond to the BRST fixed point equation (\ref{fixed point eqs}), 
which leads the same result (\ref{result of the partition function}). 

\subsection{Observables and Ward-Takahashi identities}

Let us next consider observables in this theory. 
In the context of topological field theory, such operators that are in $Q_s$-cohomology 
are called physical operators. 
In general, the physical observable has a non-trivial vev, 
while that of the $Q_s$-exact operator vanishes.
An important physical observable in our system is the sKM action introduced in the previous section.
Indeed, the sKM action (\ref{sKM action}) satisfies
\be
Q_s S_{\rm sKM} = 0,
\label{Q-closedness of sKM}
\ee
but it is not $Q_s$-exact.
The potential part of the sKM action, which is a function of $\Phi_x$ only, is apparently $Q_s$-closed because of the BRST transformation
$Q_s\Phi_x=0$. 
Although the $Q_s$-closedness of the residual part of the sKM action is not so much clear
at the first sight, we can see it by the identity, 
\be
Q_s \left[ -i \Tr \Lambda_{xy} \Phi_y U_{xy}^\dag \right] = 
-\Tr\left\{
\Phi_xU_{xy}\Phi_y U_{xy}^\dag 
-\frac{1}{2}\lambda_{xy} [U_{xy}\Phi_yU_{xy}^\dag, \lambda_{xy}]
\right\}
+\Tr \Phi_x^2,
\ee
which includes a part of the sKM action.
Noting that $Q_s^2=0$ on the gauge invariant operator and trivially $Q_s\Tr \Phi_x^2=0$,
we immediately conclude (\ref{Q-closedness of sKM}).

In addition, using the fact that the vev of the $Q_s$-exact operator vanishes, 
we find a Ward-Takahashi identity in the supersymmetric lattice gauge theory
\be
\left\langle
S_{\rm sKM}
\right\rangle
=-\left\langle
\sum_{\langle xy \rangle\in L}
\Tr \Phi_x^2
\right\rangle
+\left\langle
\sum_{x \in S}
\Tr V(\Phi_x)
\right\rangle
\ee
As we will see, we can explicitly check this identity form the localization point of view.

The sKM action is a ``good'' observable in the supersymmetric lattice gauge theory in the above sense.
So we can exactly evaluate the vev of the sKM action. In particular, the exponent of the sKM action induces
potentials of the scalar field $\Phi_x$
\be
\left\langle
e^{\gamma S_{\rm sKM}}
\right\rangle=
\left\langle
e^{\gamma \Tr \left\{
-\sum_{\langle xy \rangle\in L}
\Tr \Phi_x^2
+\sum_{x \in S}
\Tr V(\Phi_x)
\right\}}
\right\rangle,
\ee
where $\gamma$ is an arbitrary parameter and
we have flipped the sign of the coupling constant in front of the sKM action
to
utilize for a regulator of the flat directions of the supersymmetric lattice gauge theory.

Repeating the localization argument, we can evaluate the vev of the sKM model action exactly by
\be
\left\langle e^{\gamma S_{\rm sKM}}\right\rangle
=\int \prod_{x\in S} \prod_{i=1}^N \frac{d\phi_{x,i}}{2\pi i}
\frac{\prod_{x\in S}\prod_{i<j}(\phi_{x,i}-\phi_{x,j})\prod_{f\in F}\prod_{i<j}(\phi_{f,i}-\phi_{f,j})}
{\prod_{\langle xy \rangle \in L}\prod_{i\leq j}(\phi_{x,i}-\phi_{y,j})} e^{\gamma S_{\rm sKM}}.
\label{sKM integral}
\ee
The fixed points (poles) are classified by the permutation group again. For the vev of the sKM model, we see
\be
\left\langle
S_{\rm sKM}
\right\rangle
 =-\sum_{\langle xy \rangle\in L} \sum_{i=1}^N \phi_{x,i}^2+\sum_{x\in S} \sum_{i=1}^N \phi_{x,i}^2
  = (n_S-n_L) \sum_{i=1}^N \phi_{i}^2,
\ee
at the each fixed point. The measure gives the same contribution $\prod_{i<j}(\phi_{i}-\phi_{j})^{\chi}$
as the partition function. We then obtain
\be
\left\langle e^{\gamma S_{\rm sKM}}\right\rangle = {\cal C} 
 \int  \prod_{i=1}^N d\phi_{i}
\prod_{i<j}(\phi_{i}-\phi_{j})
^{\chi}
e^{\gamma(n_S-n_L) \sum_{i=1}^N \phi_{i}^2},
\label{vev of sKM}
\ee
where the number of the possible permutations (fixed points) ${\cal C}$ appears again.
Noting that $n_S-n_L = \chi-n_F$ by using the definition of the Euler characteristic,
the coefficient of the potential $\gamma(n_S-n_L)$ becomes negative for the large $n_F$,
since $\chi$ is constant for the same Riemann surface.
So the vev (\ref{vev of sKM}) is regularized in the sense of the Gaussian integral,
in comparison with the partition function itself.

Finally, we would like to discuss the continuum limit of (\ref{vev of sKM}).
Let $a^2$ denote the average area of the faces. 
As discussed in \cite{Matsuura:2014kha}, the continuum limit is defined by
$a \to 0$ and $n_F \to \infty$ with fixing the combination $a^2 n_F$ to the   
the total area of the Riemann surface ${\cal A}$. 
The scalar field $\Phi_x$ in the lattice theory is related with the  continuum field $\Phi(x)$
such that $\Phi_x = a \Phi(x)$. 
If we use the discretization of the Riemann surface with the same Euler characteristic (genus),
we find 
\be
\gamma a^2 (n_S-n_L)\sum_{i=1}^N \tilde{\phi}_{i}^2
= \gamma a^2 (\chi-n_F)\sum_{i=1}^N \tilde{\phi}_{i}^2
\  \to \   -\gamma {\cal A} \sum_{i=1}^N \tilde{\phi}_{i}^2,
\ee
where $\tilde{\phi}_{i}$'s are eigenvalues of the continuum field.
Then we obtain, in the continuum limit,
\be
\left\langle e^{\gamma S_{\rm sKM}}\right\rangle = \tilde{{\cal C}}
 \int  \prod_{i=1}^N d\tilde{\phi}_{i}
\prod_{i<j}(\tilde{\phi}_{i}-\tilde{\phi}_{j})
^{\chi}
e^{-\gamma {\cal A}\sum_{i=1}^N \tilde{\phi}_{i}^2},
\ee
where $\tilde{{\cal C}} \equiv {\cal C} a^{N+\chi N(N-1)/2}$ is also fixed.
This expression is essentially the same as the partition function of two-dimensional YM theory
appeared in (\ref{Migdal partition function}), except for the summation over the flux configurations.
Thus we successfully reproduce the perturbative partition function of the continuous two-dimensional YM theory
from the continuum limit of the discretized theory.

\section{Conclusion and Discussion}

In this paper, we discussed the localization mechanism in the various unitary matrix models,
which includes the two-dimensional supersymmetric gauge theory on 
a generic discretized Riemann surface  (generalized Sugino model).
The integrability of the unitary matrix models based on the localization 
still holds as well as the lower dimensional continuous gauge theories.

We also find that the integral formula of the partition function of the two-dimensional
supersymmetric lattice gauge theory is identical with the continuum one.
It depends only on the Euler characteristic and size of the system (topology and area of the Riemann surface).
This fact may come from the specialty of the two-dimensional YM theory, which is almost topological,
namely invariant under the area preserving diffeomorphism.
The two-dimensional discretized YM theory inherits this topological property, and so is solved exactly.
The potential gain of this study is simplification of numerical analysis of the supersymmetric lattice models.
While several numerical studies of the Sugino models have been in progress \cite{Suzuki:2007jt, Kanamori:2007ye, Kanamori:2007yx, Kanamori:2008bk,Hanada:2009hq,Hanada:2011qx,Matsuura:2014pua}, our reduced path integral would 
simplify and accelerate the numerical calculations.

This work for the first time evaluates completely the lattice path integrals by the localization technique, which
can be seen as the multi-matrix extension of the HCIZ integral based on the equivariant cohomology.
In this sense, our study connects the well-defined equivariant localization to the empirical supersymmetric localization, which backs up validity of the localization technique in the field theory.

We here frankly refer to an insufficient point of this work: 
We have discussed the partition function itself and some vevs of the physical observables
without summing up the non-perturbative flux configurations, since we do not have an operator
depending on the flux.
However, as we mentioned in the introduction, the continuum theory has a specific operator
which depends on the fluxes
and we obtain the partition function of the {\it bosonic} (non-supersymmetric)
two-dimensional YM theory as the vev of the operator.
It is an interesting problem to find the corresponding operator, which depends on the non-trivial fluxes
and reproduces the partition function of the bosonic lattice gauge theory.

We finally comment on the relation to the quiver gauge theory.
We have considered the multi unitary matrix model on the lattice,
while we can also regard it as a quiver (unitary) matrix model associated with the lattice structure:
We identify sites, links and faces with nodes, arrows and loops (superpotentials) in the quiver matrix model, respectively.
It is known that the quiver theories, including the quiver matrix models and quiver quantum mechanics,
are important in the context of the superstring (supergravity) theory or $M$-theory.
(See e.g. \cite{Denef:2002ru,Ohta:2014ria}.)
We expect that our exact result and simulation techniques in the supersymmetric lattice gauge theories
also shed light on the superstring and $M$-theory.

\section*{Acknowledgements}

The authors would like to thank
K.~Murata,
S.~Ramgoolam,
N.~Sakai,
Y.~Sasai,
F.~Sugino,
T.~Tada,
and Y.~Yoshida 
for useful discussions. 
The work of S.M., T.M. and K.O. was supported in part by Grant-in-Aid
for Young Scientists (B), 23740197,
Grant-in-Aid
for Young Scientists (B), 26800417,
and JSPS KAKENHI Grant Number 14485514, respectively.

\appendix
\section{Derivation of the 1-loop Determinants}

Here we derive the 1-loop determinant for a general matrix model induced by the supersymmetric Yang-Mills theory.
Let us first consider a set of  the bosonic matrix variables $\B^I$ and the fermionic matrix variables
$\F^I$, except for $\Phi$ which satisfies $Q\Phi=0$.\footnote{
In the supersymmetric lattice gauge theory, we have multiple $\Phi_x$'s,
but we here consider a single $\Phi$ only without loss of generality.
} We assume that
$\frac{\delta Q\F^I}{\delta \F^J}=\frac{\delta Q\B^I}{\delta \B^J}=0$.

The $Q$-exact action is
\be
\begin{split}
S &= t Q\Tr\bigg[
g_{IJ} \F^I  \overline{Q\F^J}
\bigg]\\
&=t \Tr\bigg[
||Q \vec{\F}||^2 - \F^I Q(g_{IJ} \overline{Q\F^J})
\bigg],
\end{split}
\ee
where the metric $g_{IJ}$ is a scalar function of $\vec{\B}$ only, and $|| \cdots ||^2$ denotes
a suitable norm of the vector of the fields. 
We can show that a partition function with respect to the above action is independent of the coupling $t$,
and the path integral localizes at the fixed point equation
$Q\F^I=Q\B^I=0$.

If we denote the solution
of the fixed point equation by $\B_0^I$ and $\F_0^I$, then we can
expand the fields around the fixed point by
\be
\begin{split}
\B^I &= \B_0^I + \frac{1}{\sqrt{t}}\tilde{\B}^I,\\
\F^I &= \F_0^I + \frac{1}{\sqrt{t}}\tilde{\F}^I.\\
\end{split}
\label{expansion around the fixed point}
\ee
Substituting the expansion (\ref{expansion around the fixed point}),
up to the quadratic order, the action becomes,
\be
S = \Tr\bigg[
G_{IJ} \tilde{\B}^I \tilde{\B}^J
-\Omega_{IJ} \tilde{\F}^I\tilde{\F}^J
\bigg] + {\cal O}(1/\sqrt{t}),
\label{quadratic action}
\ee
where
\be
\begin{split}
G_{IJ}
&= \left.\frac{\delta^2 }{\delta \B^I \delta \B^J}
||Q \vec{\F}||^2\right|_{\vec{\B}=\vec{\B}_0},\\
\Omega_{IJ} &=
\frac{1}{2}
\left.
\left(
\frac{\delta}{\delta \F^I} Q(g_{JK} \overline{Q\F^K})
-\frac{\delta}{\delta \F^J} Q(g_{IK} \overline{Q\F^K})
\right)
\right|_{\vec{\F}=\vec{\F}_0}
\end{split}
\ee
The quadratic action (\ref{quadratic action}) itself should be $Q$-closed (supersymmetric)
since it is independent of the coupling $t$. So we find
\be
G_{IJ}(Q\tilde{\B}^I)\tilde{\B}^J = \Omega_{IJ}(Q\tilde{\F}^I)\tilde{\F}^J,
\label{relation from Q-closedness}
\ee
where we have used the fact that $Q G_{IJ} = Q \Omega_{IJ}=0$ since $G_{IJ}$ and $\Omega_{IJ}$ are defined at the fixed point value and behave as constants.

Let us next consider an expansion of $Q\F^I$ and $Q\B^I$ around the fixed point
\be
\begin{split}
Q\F^I &= \left.Q\F^I\right|_{\vec{\B}=\vec{\B}_0}
+\frac{1}{\sqrt{t}}
\left.\frac{\delta Q\F^I}{\delta \B^J}\right|_{\vec{\B}=\vec{\B}_0}\tilde{\B}^J,\\
Q\B^I &= \left.Q\B^I\right|_{\vec{\F}=\vec{\F}_0}
+\frac{1}{\sqrt{t}}
\left.\frac{\delta Q\B^I}{\delta \F^J}\right|_{\vec{\F}=\vec{\F}_0}\tilde{\F}^J
\end{split}
\ee
while, from (\ref{expansion around the fixed point}), we see
\be
\begin{split}
Q\F^I &= Q\F_0^I
+\frac{1}{\sqrt{t}}
Q\tilde{\F}^I,\\
Q\B^I &= Q\B_0^I
+\frac{1}{\sqrt{t}}
Q\tilde{\B}^I.
\end{split}
\ee
Then we have
\be
\begin{split}
Q\tilde{\F}^I
&= \left.\frac{\delta Q\F^I}{\delta \B^J}\right|_{\vec{\B}=\vec{\B}_0}\tilde{\B}^J,\\
Q\tilde{\B}^I
&=\left.\frac{\delta Q\B^I}{\delta \F^J}\right|_{\vec{\F}=\vec{\F}_0}\tilde{\F}^J
\end{split}
\label{Q tilde}
\ee
Substituting (\ref{Q tilde}) into (\ref{relation from Q-closedness}),
we find a relation
\be
G_{IJ}\left.\frac{\delta Q\B^I}{\delta \F^K}\right|_{\vec{\F}=\vec{\F}_0}
= \Omega_{IK}\left.\frac{\delta Q\F^I}{\delta \B^J}\right|_{\vec{\B}=\vec{\B}_0}
\ee
Thus we obtain a relation between determinants of  $G_{IJ}$ and $\Omega_{IJ}$
\be
\frac{\Det \Omega_{IJ}}{\Det G_{IJ}}
= \frac{\Det \frac{\delta Q\B^I}{\delta \F^J}}{\Det \frac{\delta Q\F^I}{\delta \B^J}},
\ee
at the fixed points.

Using the above relations, we can evaluate the partition function by
\be
\begin{split}
Z &= \int \prod_I {\cal D} \B^I{\cal D} \F^I \, e^{-S(\vec{\B},\vec{\F})}\\
&=\sum_{\text{fixed points}} \int \prod_I {\cal D} \tilde{\B}^I{\cal D} \tilde{\F}^I \,
e^{-\Tr\left[G_{IJ}\tilde{\B}^I\tilde{\B}^J-\Omega_{IJ}\tilde{\F}^I\tilde{\F}^J\right]}\\
&=\sum_{\text{fixed points}} \sqrt{\frac{\Det \Omega_{IJ}}{\Det G_{IJ}}}\\
&= \sum_{\text{fixed points}} \sqrt{\frac{\Det \frac{\delta Q\B^I}{\delta \F^J}}{\Det \frac{\delta Q\F^I}{\delta \B^J}}}.
\end{split}
\ee
This is a formula of the 1-loop determinant.

%
%
%
%


\providecommand{\href}[2]{#2}
\begingroup
\raggedright

\endgroup

\end{document}